\documentclass[twoside]{article}
\usepackage{fleqn,espcrc2}
\usepackage{psfig}
\usepackage{rotate}

\newcommand{\AmS}{{\protect\the\texfont2
A\kern-.1667em\lower.5ex\hbox{M}\kern-.125emS}}
\title{Phonons in Intrinsic Josephson Systems with Parallel Magnetic Field}

\author{C. Preis\address{Institut f\"ur Theoretische Physik, Universit\"at  
Regensburg, D-93040 Regensburg, Germany}, C. Helm\address{Los Alamos
National Laboratory, T-11, Los Alamos, NM 87545, USA},
K. Schmalzl$^{\rm a}$, J.
Keller$^{\rm a}$, R. Kleiner\address{Physikalisches Institut, Universit\"at
T\"ubingen, D-72076 T\"ubingen, Germany}, P. M\"uller\address{Physikalisches
Institut III, Universit\"at Erlangen-N\"urnberg, D-91058 Erlangen, Germany}}

\begin{document}
\begin{abstract}
Subgap resonances in the $I$-$V$ curves of layered superconductors are 
explained by the coupling between Josephson oscillations and phonons with
dispersion in c-direction. In the
presence of a  magnetic field applied parallel to the layers additional 
structures due to fluxon motion appear. Their coupling with phonons  
is investigated theoretically and a shift of the phonon resonances 
in  strong magnetic fields is predicted. 

\end{abstract}

\maketitle

\section{Excitation of phonons by Josephson oscillations}
\label{sec:excit-phon-josephs}
The c-axis transport in the highly anisotropic cuprate-superconductors ${\rm
Tl_2Ba_2Ca_2Cu_3O_{10+\delta}}$ (TBCCO) and ${\rm Bi_2Sr_2CaCu_2O_{8+\delta}}$
(BSCCO) can well be described by a model where the superconducting CuO$_2$
layers are coupled by tunneling barriers forming a stack of Josephson
junctions \cite{Kleiner92,Kleiner94b}. 

Subgap structures \cite{Schl96,Yurgens96} in the $I$-$V$ characteristics could be explained by the
excitation of longitudinal c-axis phonons by the Josephson oscillations in
resistive junctions \cite{Helm97a,Helm97b,Schl98}. 
In Ref.\ \cite{Helm00} a microscopic theory for the coupling between
Josephson oscillations and phonons has been derived for the case of a short
stack of Josephson junctions where it can be assumed that the gauge-invariant
phase difference $\gamma_n(x,t)=\varphi_n(x,t) - \varphi_{n+1}(x,t) -
\frac{2e}{\hbar} \int_n^{n+1}dz A_z(x,z,t)$ between superconducting layers $n$
and $n+1$ is constant along the layers (independent of $x$). In the simplest
model the tunneling current through a junction $n$  is relatexd the bias current 
density $j_b$ by the RSJ equation   
\begin{equation} 
  j_b=j_c\sin \gamma_n(t) + \sigma E_n(t) + \dot D_n(t)
  \label{RSJ} 
.
\end{equation}
Here $E_n(t)$ is the average electric field across the junction which is
related to the phase difference $\gamma_n$ by the second Josephson equation
$\hbar \dot \gamma_n=2ed E_n$ where $d$ is the thickness of the barrier. 

$D_n=E^\rho_n$ is the field generated by the oscillating conduction 
electron charges on the CuO$_2$ layers. It can be expressed by the average
field in the barrier and the ionic polarization, $D_n= \epsilon_0E_n + P_n$. 
The polarization $P_n$ is proportional to the lattice displacements of
ions. Phonons are excited by the field of the oscillating electronic charges
on the CuO$_2$ layers. The field $D(k_z)$  can be expressed by a
generalized dielectric function $D(k_z) = \epsilon_0\epsilon^{\rm
ph}_{zz}(k_z,\omega) E(k_z)$ of the form 
\begin{equation}
 \epsilon_{zz}^{\rm ph}(k_z,\omega) = \Bigl(1-
 \sum_\lambda
  \frac{\vert\Omega(k_z,\lambda)\vert^2 } {\omega^2(k_z,\lambda)-\omega^2}
\Bigr)^{-1}.
\end{equation}

Here $\omega(k_z,\lambda)$ are the eigenfrequencies of the dynamical matrix
(including long-range Coulomb forces). The oscillator strength $\vert
\Omega(k_z,\lambda)\vert^2$ takes care of the fact that ions inside the
barrier and on the CuO$_2$ layers are excited by different fields and 
have to be counted differently in the polarization in the 
RSJ-equation (\ref{RSJ}).

The dielectric function has zeros at the frequencies of
longitudinal c-axis phonons. It can also be written in the more common form
\begin{equation}
  \epsilon_{zz}^{\rm ph}(k_z,\omega)= \epsilon_\infty + 
  \sum_\lambda \frac{\vert\tilde \Omega(k_z,\lambda)\vert^2 }
  {\tilde \omega^2(k_z,\lambda)-\omega^2} 
\end{equation}
where we have included a background DK.
The frequencies $\tilde \omega(k_z,\lambda)$ where the dielectric function
has poles can also be calculated from a dynamical matrix where long-range
Coulomb forces have been subtracted. In the limit $k_z \to 0$ they 
correspond to the transversal optical eigenfrequencies of the system. 

In the resistive state the phase difference of barrier $n$ has the form
\begin{equation}
  \gamma_n(t) = \theta_n + \omega t + \delta \gamma_n(t)
\end{equation}
where $\delta \gamma_n(t)$ oscillates with the same Josephson frequency
$\omega$, while for a barrier in the superconducting state the term $\omega t$
is missing. Inserting this ansatz in the RSJ equations we can solve for the dc
part and the oscillating part.  
In the special case where only one junction of the stack 
is in the resistive state we obtain for
the normalized dc-current $J_b:=j_b/j_c$ as function of the dc-voltage
$V=\hbar\omega/(2e)$:
\begin{equation}
  J_b= J_{qp}(V) + \frac{\omega_J^2}{2\omega^2} 
  \frac{{\bar \epsilon}_2
    +\sigma/(\epsilon_0\omega)}{\bar\epsilon^2_1 + (\bar \epsilon_2 +
    \sigma/(\epsilon_0\omega))^2} 
.\end{equation}
Here $\bar \epsilon(\omega)=\bar\epsilon_1(\omega)+ i\bar \epsilon_2(\omega)$
is an averaged phonon dielectric function defined by 
\begin{equation}
  \bar\epsilon(\omega) = 
G^{-1}(\omega)
+ 
  \frac{\bar\omega_J^2}{\omega^2} - \frac{i\sigma}{\epsilon_0\omega}
,
\end{equation}
\begin{equation}
  G(\omega)=
  \frac{1}{N_z}\sum_{k_z} 
  \frac{1}{\epsilon^{\rm ph}_{zz}(k_z,\omega) -
    \frac{\bar\omega_J^2}{\omega^2} +
  \frac{i\sigma}{\epsilon_0\omega}}  
.
\end{equation}
Here $\bar\omega^2_J= \omega^2_J\sqrt{1-(j/j_c)^2}$ and $\omega_J^2=
2edj_c/(\hbar \epsilon_0)$ is the bare Josephson plasma frequency. 
In the case of a
dispersionless phonon we just have $\bar \epsilon(\omega) = \epsilon^{\rm
ph}_{zz}(\omega)$. 

In the general case the $I$-$V$ curve shows peaks at the zeros of the real part of
$\bar\epsilon(\omega)$ corresponding to c-axis phonons with a high density of
states and non-vanishing oscillator strength $\vert \Omega\vert^2$. Besides
optical $\Gamma$-point phonons also phonons from the edge of the Brillouin zone
at $k_z=\pi/d$ and even acoustical phonons may contribute to phonon resonances
in the $I$-$V$ curves \cite{Helm00}.

The coupling to an acoustical phonon at $k_z=\pi/d$ may explain a resonance
observed for TBCCO at 3.2 mV by Seidel et al. \cite{Seidel97} at a
frequency/voltage which is lower then any optical phonon branch expected from
model calculations. A fit to the experimental data can be made with reasonable
values of the oscillator strength and phonon damping which are compatible with
optical experiments. Also a double peak structure found in BSCCO \cite{Schl98}
may be due to the coupling to one  optical branch with two van Hove
singularities at $k_z=0$ and $k_z=\pi/d$.

The dispersion of phonons also leads to a coupling (phase-locking) of 
Josephson oscillations in different resistive junctions
\cite{Helm00,Preis00}. 
A phase-locking in a stack of Josephson junctions is important for applications
of such systems  as high-frequency mixers and detectors.
 
\section{Vortex motion and phonons}

In the case of long junctions and in the
presence of an external magnetic field applied parallel to the layers we have
to account for the variation of the phase $\gamma_n(x,t)$ along the
layers (x-direction). The result is a generalization of the coupled sine-Gordon
equations derived in  \cite{Sakai93,Kleiner94,Bulaevskii94}: 
\begin{equation}
\partial^2_x \gamma_n(x,t)= \frac{1}{\lambda_J^2} J_n -
\frac{1}{\lambda_K^2}(J_{n+1} + J_{n-1}) \label{sineGordon}
\end{equation}
\begin{equation}
J_n = \sin \gamma_n + \frac{\sigma}{\epsilon_0\omega_J^2} \dot \gamma_n +
\frac{1}{\omega_J^2}\epsilon^{\rm ph}_{zz}\ddot \gamma_n.
\end{equation}
The characteristic lengths for the variation of the phase along the layers 
are calculated from 
$$
\frac{1}{\lambda_K^2}= \frac{d^2}{\lambda_c^2\lambda^2_{ab}};\quad
\frac{1}{\lambda_J^2}=\frac{1}{\lambda_c^2}+\frac{2}{\lambda_K^2}
$$
where $\lambda_c$, $\lambda_{ab}$ are the penetration depths of  magnetic
fields polarized in c-direction and parallel to the ab-planes,
respectively. They are related to the corresponding plasma frequencies
$\lambda_c = c/\omega_J$, $\lambda_{ab}= c/\omega_{ab}$, where $c$ is the
velocity of light. For BSCCO realistic values are $\lambda_{ab}=170$nm,
$d=1.5$nm, $\lambda_c=150\mu$m, which gives a Josephson penetration depth of 
$\lambda_J\simeq 1 \mu$m.
  
In the derivation of Eq. (\ref{sineGordon}) terms of the form $\ddot
\gamma/\omega^2_{ab}$, which are small compared to $\gamma$ for the
frequency range considered, have been neglected. Furthermore in the
coupling to phonons only polarization effects with polarization in
c-direction excited by fields in c-direction (expressed by the
dielectric function $\epsilon^{\rm ph}_{zz}(k_x,k_z,\omega)$) have been
taken into account. It can be shown \cite{Preis00}, that the neglect of  other
polarizations is a good approximation for wave-vectors with $k_xd \ll 1$,
which is well fulfilled for $k_x \le 1/\lambda_J$. On the other hand,
$k_z$ is not small and $k_z\gg k_x$ in general.  

In the following numerical calculations we consider only
local polarization effects, i.e. we neglect the dispersion of phonons. This
will be sufficient to demonstrate the principle effects.

\subsection{Boundary conditions}

Assuming a stack with $N$ Josephson barriers with $n=1 \dots N+1$
superconducting layers and two normal contacts $n=0, N+2$ this set of
equations (\ref{sineGordon}) holds for $n=1 \dots N$. In the barriers
connecting the superconducting layers with the normal contact $J_n$ has to be
replaced by the normalized bias current $J_b= j_b/j_c$. It is useful to
incorporate this boundary condition into the set of equations by subtracting
$J_b$ for each term writing $\tilde J_n:= J_n-J_b$ then 
\begin{equation}
  \partial^2_x\gamma_n - \frac{1}{\lambda_c^2} J_b =
  \frac{1}{\lambda_J^2}\tilde J_n - \frac{1}{\lambda_K^2}\Bigl( \tilde
  J_{n+1} + \tilde J_{n-1}\Bigr) 
  \label{sineGordon2}
\end{equation}
which has to be solved  with the boundary conditions $\tilde J_n=0$ for $n=0,
N+1$ for a finite stack. 

The  magnetic field (in y-direction) which causes the variation of the phase
along the layers  enters explicitly the boundary condition for $\partial_x
\gamma_n$ at the edges of the
stack. It consists of the external field $B_{ext}$ and the field generated by
the bias current. If we neglect the latter we have 
\begin{equation}
  \partial_x\gamma_n(x=0)=\partial_x \gamma_n(x=L_x)=
 \frac{2ed}{\hbar}  B_{ext}=:\eta
\end{equation}
and we may drop the small contribution $J_b/\lambda_c^2$ on the left
side of Eq. (\ref{sineGordon2}). Then in the case of a constant phase along
the layers one recovers the RSJ equation $\tilde J_n=0$.

\subsection{Analytical solution}

The set of coupled sine-Gordon equations can be solved numerically. Here one
makes the general ansatz for the phases in the different layers:
\begin{equation}
  \gamma_n(x,t) =\Gamma_n(t) + \eta x 
  +\sum_{m=1}^M\delta \gamma_n(m,t) \cos(\frac{m \pi x}{L_x})
  \label{gammax}
\end{equation}
using a Fourier expansion for the spatially oscillating part of the phases
and splitting off the term $\eta x$ which takes care of the average
increase of the phase difference due to the applied magnetic field. With help
of this ansatz a coupled set of differential equations for the components
$\delta \gamma_n(m,t)$ can be derived and solved with Runge-Kutta techniques.

Approximate analytical solutions are possible by setting 
\begin{equation}
  \Gamma_n(t) = \theta_n + \omega t + \delta\gamma_n(0,t)
\end{equation}
and linearizing the equations (\ref{sineGordon})  with 
respect to small oscillating terms $\delta\gamma_n(m,t)$. This ansatz
is well justified for large magnetic fields where the magnetic flux 
penetrates the stack almost homogeneously. 
Therefore there is a voltage drop 
$V_n=\langle\dot \gamma_n(t)\rangle\hbar/(2e) = \omega\hbar/(2e)$ 
over each junction.  The
oscillating part describes standing waves which oscillate primarily with the
same basic Josephson frequency $\omega$.

Using in addition a Fourier expansion of the oscillating parts in c-direction
and keeping only the lowest harmonics in $\omega$ we arrive
at the following expression for the dc-current 
\begin{equation}
J_b =  \frac{\omega\sigma}{\epsilon_0\omega_{\rm J}^2}
-  2\frac{\omega_{\rm J}^2}{\omega^2}
  \frac{1}{N^2}\sum_{k_x,k_z} {\rm Im}( P(k_x,k_z,\omega))
\end{equation}
with
$$
P(k_x,k_z,\omega)= 
  {\frac{\vert I(k_x,\eta)\vert^2
      \vert p(k_z)\vert^2}{
      \epsilon^{\rm ph}_{zz}(k_x,k_z,\omega)+
      i\frac{\displaystyle \sigma}{\displaystyle \epsilon_0\omega} 
-\frac{\displaystyle k_x^2{{\tilde c}^2(k_z)}}{\displaystyle\omega^2} 
}}
$$ 
where the sum goes over  the discrete values of $k_x=m \pi/L_x$,
$k_z=n\pi/((N+1)d)$, $ 0<|n|\leq N$.

The denominator contains the phonon dielectric function and the
characteristic velocity
\begin{equation}
\tilde c(k_z)= c/\sqrt{1- 2 \frac{\lambda_{ab}^2}{d^2} (\cos(k_z d) - 1)}
.
\end{equation}
Resonances are expected at frequencies where the real part of the
denominator vanishes. There size depends on the weighting functions 
\begin{equation} 
p(k_z)=\sum_n e^{-i\theta_n}e^{-ik_zn}
\end{equation}
which contains a Fourier transformation of the static phase distribution
and the function 
\begin{equation}
I(k_x,\eta)=\frac{i}{2L_x}\int_0^{L_x}{dx}
  e^{-i\eta x}e^{ik_x x}
\end{equation}
depending on the magnetic field, which is peaked at $k_x\simeq\eta$.

\section{Single contact in a magnetic field}

For a single contact, $N=1$, the value of $k_z$ is fixed to $k_z=\pi/(2d)$.
Then the characteristic (Swihart-) velocity  is 
$\tilde c= c_0/\sqrt{1+2\lambda^2_{ab}/d^2}=\omega_J\lambda_J$, and
the resonance frequencies in the $I$-$V$ curve are determined from  
\begin{equation} 
  {\rm Re}\Bigl[ \epsilon^{\rm ph}_{zz}(\omega)  - 
  (\frac{k_x\tilde c}{\omega})^2\Bigr]=0.  
\end{equation} 
For a constant $\epsilon^{\rm ph}_{zz}$ structures appear in the
current-voltage characteristic at $\omega_{res,m}= m
\pi\tilde c/(\sqrt{\epsilon^{\rm ph}_{zz}}L_x)$. These are the well-known Fiske
steps \cite{Fiske64}. They correspond to the excitation of standing
electromagnetic waves in the Josephson junction of length $L_x$. The largest 
amplitude is  obtained for wave-vectors $k_x\simeq \eta$. Here the
velocity of fluxons equals  the phase velocity of the electromagnetic waves. 
In the case of a very long junction the Fiske steps merge
into one flux flow branch which is peaked at 
$V =\hbar \omega_{\rm res}/(2e) = \tilde c d B_{\rm ext}/\sqrt{
  \epsilon^{\rm ph}_{zz}}$
(Eck-peak \cite{Eck64}). 

In order to discuss the influence of phonons we use 
here for simplicity a dispersionless optical phonon band with the
dielectric function  
\begin{equation} 
  \epsilon^{\rm ph}_{zz}(\omega) =
  \epsilon_\infty + \frac{\vert \Omega\vert^2}{\omega^2_{\rm TO} - 
    \omega^2 - i r\omega} 
.
\end{equation}  
The spectrum of resonances as function of the discrete values of
$k_x=m\pi/L_x$ is shown in Fig.\ 1. One obtains two branches: for small
$k_x$-values the lower branch corresponds to the propagation of
electromagnetic waves, while the upper branch is phonon-like. The
lower branch ends
at the (transverse) phonon eigenfrequency $\omega_{\rm TO}$.
The upper branch starts at the zero of the dielectric function, i.e. at the
longitudinal eigenfrequency $\omega_{\rm LO}$. The parameters used to
calculate the dispersion shown in Fig.\ 1 are adapted to TBCCO. 

In Fig.\ 2 the result for the
$I$-$V$ curve is shown for three different magnetic fields. The figures compare
numerical (Runge-Kutta)  with analytical results. Note that the magnetic
field selects the $k_x$-value where the strongest resonance occurs: 
$k^{\rm max}_x=\eta = (2ed/\hbar)B_{\rm ext}$. 
In order to show the phonon effects
more clearly a large separation between $\omega_{\rm TO}$ and $\omega_{\rm
LO}$ has been chosen. Furthermore for numerical reasons a comparatively small
value of the McCumber parameter $\beta_c=50$ has been used, while
$\beta_c=500$ would be more realistic ($\beta_c=\omega_c^2/\omega_J^2$ with
$\omega_c= 2edj_c/(\hbar\sigma)$).     
\begin{figure}
  \psfig{figure=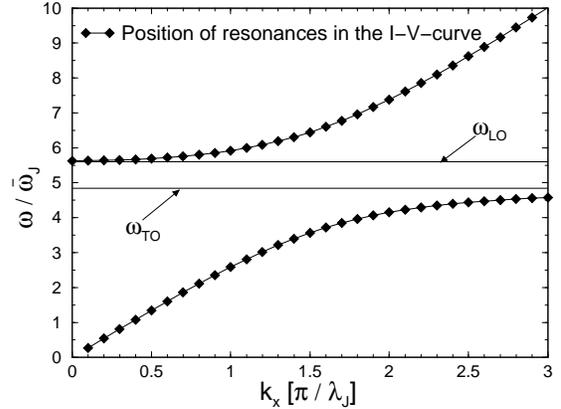,width=0.45\textwidth} 
  \caption{Resonance frequencies for a single Josephson contact
    with one dispersionless phonon mode.} 
\end{figure}
For $B_{\rm ext}=0$ only one resonance occurs at  $\omega_{\rm LO}$
corresponding to the subgap-resonance discussed in Sec.\
\ref{sec:excit-phon-josephs}.
For finite $B_{\rm ext}$ one finds two groups of resonances
corresponding to the two
branches in Fig.\ 1.  The fine-structure is due to Fiske-resonances in
the stack of finite length $L_x$.
With increasing field strength the upper peak shifts to higher
frequencies. The lower peak approaches the TO-frequency while 
loosing weight.  
In all cases there is a gap with no resonances for frequencies between
$\omega_{\rm TO}$ and $\omega_{\rm LO}$. 

\begin{figure}[t]
\psfig{figure=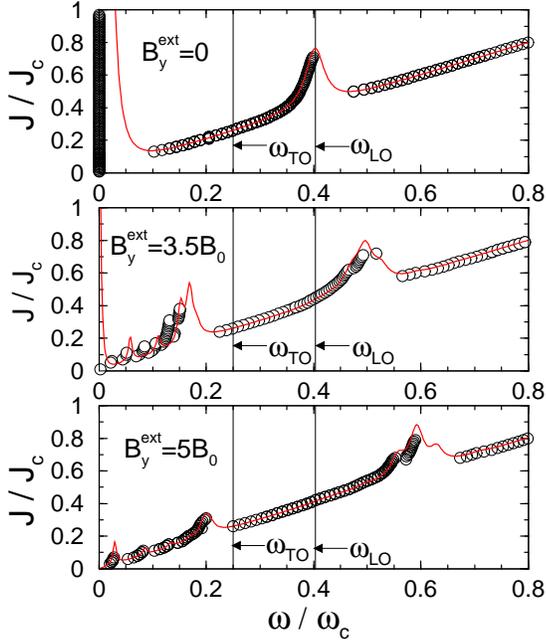,width=0.45\textwidth}
\caption{$I$-$V$ curves for a single contact for three different
  magnetic fields. $B_0=\Phi_0/(dL_x)$ corresponds to a magnetic field
  with one flux quantum per junction.} 
\end{figure}

Generally the agreement between numerical and analytical
calculations is good, in particular, concerning the position of peaks. The
agreement in the height of the peaks can be improved by going beyond the
linear approximation in the oscillation amplitudes. The numerical 
calculations show the same hysteretic behavior as the experimental
results if the bias current density is changed. 

\section{Several contacts in a magnetic field}

Numerical calculations can also be performed for larger stacks. Analytical
calculations are no longer possible for $N>2$ without further assumptions on
the relative phases $\theta_n$ for the different junctions. In Fig.\ 3  we show
results for the positions of the resonances for a stack with $N=4$ junctions
for all possible values of $k_z=n\pi/(5d)$. The multiple branches correspond
to different values of the characteristic 
velocity $\tilde c$ which depends strongly on $k_z$.

\begin{figure}
\psfig{figure=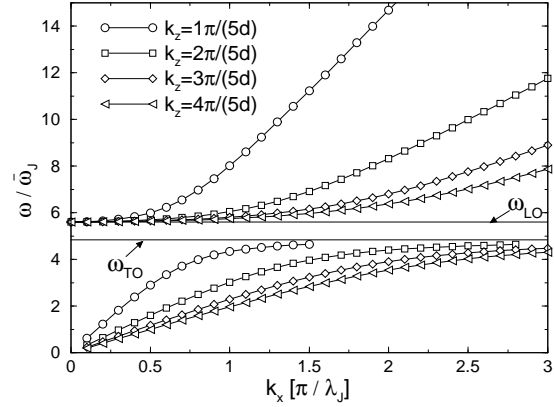,width=0.45\textwidth}
\caption{Resonance frequencies for a stack of 4 contacts.} 
\end{figure}

In Fig.\ 4  we show results for the $I$-$V$ curve for a large applied magnetic
field. Here we find a vortex lattice which is moving due to
the bias current. Both numerical and analytical results are shown as
function of frequency which  corresponds to the dc-voltage drop over
a single junction.
For our analytical calculations of the $I$-$V$
curves we have superimposed results calculated with all possible values of
$k_z$. The comparison with numerical results  in Fig.\ 4 demonstrates
that  modes
with all possible $k_z$-values will be excited by cycling the bias current.
The group of peaks at the highest frequency correspond to electromagnetic
modes with the smallest $k_z$ value. The middle group contains phonon like
excitations while the lowest group are electromagnetic excitations  of Kleiner
modes. Again a gap appears between $\omega_{\rm TO}$ and $\omega_{\rm LO}$ in
accordance with the dispersion curves in Fig.\ 3. 
\begin{figure}
\psfig{figure=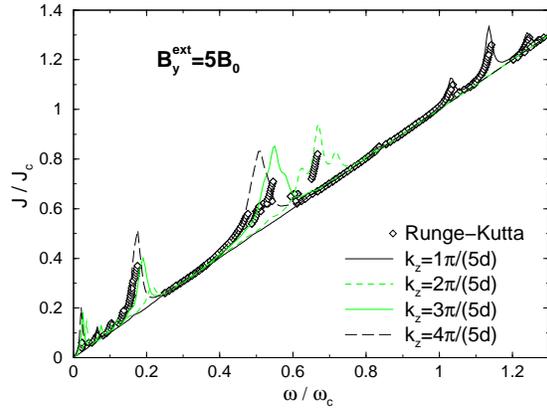,width=0.45\textwidth}
\caption{$I$-$V$ curves for a stack of 4 contacts in high magnetic field.} 
\end{figure}

\section{Comparison with experimental results and conclusions} 

The main results of the preceding section are: a) a
shift of phonon resonances in a parallel magnetic field, 
b) Fiske resonances in the frequency range
of phonons are no longer equidistant, c) the flux-flow voltage is no longer
proportional to the magnetic field and has a gap between $\omega_{\rm
  TO}$ and $\omega_{\rm LO}$. In order to observe these effects
experimentally a strong magnetic field has to be applied ($>3$T). 
In most experiments \cite{Schl98} the applied field has been much lower, and
no shift of phonon resonances has been observed.
A further requirement is that the field is applied strictly parallel to the
layers in order to avoid vortex pinning from inhomogeneities. 
The fact that in the cited experiments \cite{Schl98} neither Fiske resonances
in the frequency range of phonons nor a flux flow branch could  be observed
is an indication of vortex pinning. 
Recently the flux flow voltage has been measured for a stack of 30 junctions 
in BSCCO \cite{Lat00}. Deviations from the linear field dependence together with
anomalies in the frequency range of an optical phonon have been observed,
which supports the present model for the interaction between flux flow and phonons. 

The author would like to thank K. Schlenga and L. Bulaevskii for fruitful
discussions. Financial support by the Bayerische Forschungs\-stiftung (C.P.) and the
Department of Energy under contract W-7405-ENG-36
(C.H.) is gratefully acknowledged.

\end{document}